\documentclass{article}
\usepackage{bm}
\begin{document}
\title{$\mu-\tau$ Symmetry, Nonzero $\theta_{13}$, and CP Violation}
\author{\bf{Asan Damanik}\footnote{E-mal: d.asan@lycos.com}\\Faculty of Science and Technology\\Sanata Dharma University\\Kampus III USD Paingan Maguwoharjo Sleman Yogyakarta\\Indonesia}
\date{}

\maketitle

\begin{abstract}
If we impose the $\mu-\tau$ symmetry, as a constraint into neutrino mass matrix, one find that the Jarlskog rephasing invariant: $J_{\rm CP}=0$  which implies that CP violation cannot be accommodated in the $\mu-\tau$ symmetry scheme.  By introducing a small parameter $x$ that perturb the neutrino mass matrix with $\mu-\tau$ symmetry with trace of neutrino mass matrix remain constant, we can obtain $J_{CP}\neq 0$ and consequently $\theta_{13}\neq 0$.
\end{abstract}

\section{Introduction}
Recent experimental results as reported in \cite{Minos, Double, T2K, Daya, RENO} show that the mixing angle $\theta_{13}\neq 0$ and its value relatively large.  Nonzero $\theta_{13}$ have some implications in our understanding about the neutrino sector beyond the standard model, and one of the implications is the possibility of CP violation in neutrino sector as well as in the quarks sector.   From the theoretical side, it is possible that mixing angle $\theta_{13}\neq 0$ as one can read from the general formulation of the neutrino mixing matrix $V$  in standard parametrization:
\begin{eqnarray}
V=\bordermatrix{& & &\cr
&c_{12}c_{13} &s_{12}c_{13} &s_{13}e^{-i\delta}\cr
&-s_{12}c_{23}-c_{12}s_{23}s_{13}e^{i\delta} &c_{12}c_{23}-s_{12} s_{23}s_{13}e^{i\delta}&s_{23}c_{13}\cr
&s_{12}s_{23}-c_{12}c_{23}s_{13}e^{i\delta} &-c_{12}s_{23}-s_{12}c_{23}s_{13}e^{i\delta} &c_{23}c_{13}}
 \label{V1}
\end{eqnarray}
where $c_{ij}$ is the $\cos\theta_{ij}$, $s_{ij}$ is the $\sin\theta_{ij}$, and $\theta_{ij}$ are the mixing angles and $i,j=1,2,3$.

To explain the evidence of nonzero and relatively large $\theta_{13}$, several authors have already proposed some models.   The simple way to accommodate a nonzero $\theta_{13}$ is to modify the neutrino mixing matrix by introducing a perturbation matrix into known mixing matrix such that it can produces a nonzero $\theta_{13}$ \cite{Boudjemaa, He11, Damanik, Damanik1, RodejohanW}, and the other is to build the model by using some discrete symmetries \cite{Luca}.  The nonzero $\theta_{13}$ is also known related to the Dirac phase $\delta$ as one can see in the standard parameterization of the neutrino mixing matrix.  Thus, nonzero $\theta_{13}$ gives a clue to the possible determination of CP violation in neutrino sector. Perturbation of neutrino mixing matrix in order to accommodate both nonzero $\theta_{13}$ and CP violation have been reported \cite{HarrisonScott, Harrison1, Grimus, Friedberg, Xing, He2, Zhou}.

One of the well-known neutrino mixing matrix is tribimaximal (TBM) neutrino mixing matrix because it can predicts and explains the experimental data very well before the evidence of nonzero $\theta_{13}$ reported by some collaborations.  TBM is also very useful to obtain the neutrino mass matrix $M_{\nu}$.  The neutrino mass matrix deduced from TBM is the neutrino mass matrix with $\mu-\tau$ symmetry.  Considering from the number of parameters in the neutrino mass matrix, the $\mu-\tau$ symmetry has an advantage in reducing the number of parameters in neutrino mass matrix from 6 to be 4 parameters.

Concerning the $\mu-\tau$  symmetry and mixing angle $\theta_{13}$, Mohapatra \cite{Mohapatra} stated explicitly that neutrino mass matrix that obey $\mu-\tau$ symmetry to be the reason for maximal $\mu-\tau$ mixing, one gets $\theta_{13}=0$ and conversely if $\theta_{13}\neq 0$ can provide the $\mu-\tau$ symmetry beraking manifests in the case of normal hierarchy.  The nonzero $\theta_{13}$ and its implication to the leptogenesis as an origin of matter is discussed in \cite{Mohapatra1}.   Aizawa and Yasue \cite{Aizawa} analysis complex neutrino mass texture and the $\mu-\tau$ symmetry which can yield small $\theta_{13}$ as a $\mu-\tau$ breaking effect.   The $\mu-\tau$ symmetry breaking effect in relation with the small $\theta_{13}$ also discussed in \cite{Fuki}.  Another scenario that can produced Dirac CP phase by studying other mixing scenarios which deviate from tri-bimaximal mixing by leaving only one of the columns or one of the rows invariant which is called as "generalized trimaximal mixing" \cite{Albright, Albright1}.  Analysis of the correlation between CP violation and the $\mu-\tau$ symmetry breaking can be read in \cite{Mohapatra2, Baba, He}.  In \cite{Ge, Ge1}, the Dirac CP phace $\delta$ can be obtained by exploring the generalized $\bf {Z_{2}^{s}}$ symmetry without assumming $\mu-\tau$ symmetry, and in \cite{He1} propose the octahedral symmetry with geometry breaking can predict $\theta_{13}\neq 0$ and CP violation.

In this talk we derive nonzero $\theta_{13}$ by modifying the neutrino mass matrix constrained by $\mu-\tau$ symmetry by introducing a small parameter to perturb the neutrino mass matrix with $\mu-\tau$ symmetry.  This talk is organized as follow: in section 2, we modify neutrino mass matrix with $\mu-\tau$ symmetry.  In section 3, we determine the Dirac neutrino phase $\delta$ from perturbed neutrino mass matrix and $\theta_{13}$ qualitatively.  Finally, section 4 is devoted to conclusion.

\section{Perturbed $\mu-\tau$ symmetry and Dirac phase $\delta$}

The first neutrino mass matrix with $\mu-\tau$ symmetry pattern was derived phenomenologically by Fukuyama and Nishiura \cite{Fukuyama}.  Concerning the neutrino mass matrix constrained by $\mu-\tau$ symmetry, most of the authors have showed that the $\mu-\tau$ symmetry lead to $\theta_{13}=0$ \cite{Damanik14}.  The effect of $\mu-\tau$ symmetry broken in the neutrino mass matrix can arises the leptonic CP violation was proposed by Mohaptara and Rodejohann \cite{Mohapatra2}.  Neutrino mass matrix with $\mu-\tau$ symmetry read:
\begin{equation}
M_{\nu}=\bordermatrix{& & &\cr
&P &Q &Q\cr
&Q &R &S\cr
&Q &S &R}. \label{M}
\end{equation}

In this talk, to break the $\mu-\tau$ symmetry softly, we introduce a small paramater $x$ ($x<<R$) that perturb the neutrino mass matrix in Eq. (\ref{M}) but maintain the trace of the perturbed neutrino mass matrix remain constant.   This perturbation technique have been successfully applied by Damanik \cite{Damanik13} to perturbed neutrino mass matrix which is constrained by invariant under a cyclic permutation.  In this perturbation scenario, the neutrino mass matrix in Eq. (\ref{M}) reads:
\begin{equation}
M_{\nu}=\bordermatrix{& & &\cr
&P &Q &Q\cr
&Q &R-ix &S\cr
&Q &S &R+ix},\label{Mv2}
\end{equation}
which trace of perturbed $M_{\nu}$ is remain constant: $2R+P$.  The perturbed neutrino mass matrix in Eq. (\ref{Mv2}) produces:
\begin{equation}
M_{\nu}^{'}=\bordermatrix{& & &\cr
&a &b &c\cr
&b^{*} &d &e\cr
&c^{*} &e^{*} &d},\label{Mv3}
\end{equation}
where:
\begin{equation}
a=\left(M_{\nu}M_{\nu}^{\dagger}\right)_{ee}=\frac{2m_{1}^{2}+m_{2}^{2}}{3},\label{a}
\end{equation}
\begin{equation}
b=\left(M_{\nu}M_{\nu}^{\dagger}\right)_{e\mu}=\frac{m_{2}^{2}-m_{1}^{2}+ix(m_{2}-m_{1})}{3},\label{b}
\end{equation}
\begin{equation}
c=\left(M_{\nu}M_{\nu}^{\dagger}\right)_{e\tau}=\frac{m_{2}^{2}-m_{1}^{2}-ix(m_{2}-m_{1})}{3},\label{c}
\end{equation}
\begin{equation}
d=\left(M_{\nu}M_{\nu}^{\dagger}\right)_{\mu\mu}=\frac{m_{1}^{2}+5m_{2}^{2}}{6}+x^{2},\label{d}
\end{equation}
\begin{equation}
e=\left(M_{\nu}M_{\nu}^{\dagger}\right)_{\mu\tau}=\frac{m_{1}^{2}+m_{2}^{2}}{6}+ix\left(m_{2}(\cos(2\delta)-\frac{2}{3})-\frac{m_{1}}{3}\right).\label{e}
\end{equation}

\section{Jarlskog rephasing invariant}
The Jarlskog invariant $J_{\rm CP}$, which is very uselful for quantifying CP violation, is given by \cite{Jarlskog}:
\begin{equation} 
J_{\rm CP}={\rm Im}\left[V_{11}V_{22}V_{12}^{*}V_{21}^{*}\right].\label{C}
\end{equation}
If we insert the corresponding values of $V$ of Eq. (\ref{V1}) into Eq. (\ref{C}), then we have:
\begin{equation}
J_{\rm CP}=\frac{1}{8}\cos\theta_{13}\sin2\theta_{12}\sin2\theta_{23}\sin2\theta_{13}\sin\delta.
\end{equation}

Alternatively, the Jarlskog rephasing invariant $J_{\rm CP}$  can  be also determined from relation \cite{Branco}:
\begin{equation}
J_{\rm CP}=-\frac{{\rm Im}\left[(M_{\nu}^{'})_{e\mu}(M_{\nu}^{'})_{\mu\tau}(M_{\nu}^{'})_{\tau e}\right]}{\Delta m_{21}^{2}\Delta m_{32}^{2}\Delta m_{31}^{2}}.\label{J}
\end{equation}
where $(M_{\nu}^{'})_{ij}=(M_{\nu}M_{\nu}^{\dagger})_{ij}$ and $ i,j=e,\nu,\tau$.  For the neutrino mass matrix constrained by $\mu-\tau$ symmetry, the value of $J_{\rm CP}=0$.  But, perturbed $\mu-\tau$ symmetry as shown in Eq. (\ref{Mv2}) can give $J_{CP}\neq 0$.

Substituting Eqs. (\ref{b}), (\ref{e}), and complex conjugate of Eq. (\ref{c}) into Eq. (\ref{J}), we have the Jarlskog invariant:
\begin{equation}
J_{\rm CP}=\frac{1}{9}\frac{\left[m_{2}(\Delta m_{21}^{2})^{2}(1-A)x+(m_{2}(m_{2}-m_{1})^{2}A-\frac{1}{3}m_{1}^{3}-\frac{2}{3}m_{2}^{3}+m_{1}m_{2}^{2})x^{3}\right]}{\Delta m_{21}^{2}\Delta m_{32}^{2}\Delta m_{31}^{2}},\label{f}
\end{equation}
where $A=\cos(2\delta)$.

Because the value of the parameter $x$ is very small, the terms that related to $x^{3}$ in Eq. (\ref{f}) can be neglected and the Jarlskog invariant in this scheme as follow:
\begin{equation}
J_{\rm CP}\approx \frac{1}{9}\frac{xm_{2}\Delta m_{21}^{2}(1-A)}{\Delta m_{32}^{2}\Delta m_{31}^{2}}. \label{JJ}
\end{equation}
\section{Conclusions}
Neutrino mass matrix with $\mu-\tau$ symmetry leads to $J_{CP}=0$.  Perturbed $\mu-\tau$ symmetry neutrino mass matrix with trace of neutrino mass matrix remain constant can be used as alternatively to produce $J_{CP}\neq 0$ which also implies that mixing angle $\theta_{13}\neq 0$.  From Eqs. (\ref{J}) and (\ref{JJ}) one can determine the parameter $x$ when the value of $m_{2}$ is known.

\section*{Acknowledgments}

Author thank to the 16th Lomonosov Conference on Elementary Particle Physics Organizer for hospitality during the conference, DP2M Dikti Kemendikbud and Sanata Dharma University for financial support.

\end{document}